\documentclass[12pt]{article}
\usepackage[dvips]{graphicx} 
\usepackage{amsmath}
\usepackage{tabularx} 
\usepackage{setspace}   
\numberwithin{equation}{section}
\newcommand{\ma}{\mathcal}   
\usepackage{amsfonts} 
\usepackage{amssymb} 

\newtheorem{thm}{Theorem}
\newtheorem{cor}[thm]{Corollary}
\newtheorem{lemma}[thm]{Lemma}

\newtheorem{defn}[thm]{Definition}
\newtheorem{rmk}[thm]{Remark}

\def\be{\begin{eqnarray}}
\def\ee{\end{eqnarray}}
\def\bee{\begin{eqnarray*}}
\def\eee{\end{eqnarray*}}
\def\bmx{\begin{pmatrix}}
\def\emx{\end{pmatrix}}
\begin{document}

\title{An application of decomposable maps in proving
multiplicativity of low dimensional maps}
\author{Motohisa Fukuda\\
Department of Mathematics,\\
University of California, Davis}

\maketitle 

\abstract{
In this paper we present a class of maps for which 
the multiplicativity of the maximal output $p$-norm holds 
for $p=2$ and $p\geq 4$. 
This result is a slight generalization of 
the corresponding result in \cite{KingKoldan06}.
The class includes all positive trace-preserving maps from 
$\ma{B}(\mathbb{C}^3)$ to $\ma{B}(\mathbb{C}^2)$. 
Interestingly, by contrast, 
the multiplicativity of $p$-norm was investigated in the context of quantum information theory and
shown not to hold in general for high dimensional quantum channels \cite{HaydenWinter08}.
Moreover, the Werner-Holevo channel, 
which is a map from $\ma{B}(\mathbb{C}^3)$ to $\ma{B}(\mathbb{C}^3)$,
is a counterexample for $p>4.79$.}

\section{Introduction}
Suppose we have a map
\be
\Phi: \ma{B}(\mathbb{C}^m) \rightarrow \ma{B}(\mathbb{C}^n),
\ee
where $\ma{B}(\mathbb{C}^d)$ is the set of
(bounded) linear operators on $\mathbb{C}^d$.
Then, the maximal output $p$-norm is defined as
\be \label{MOPN}
\nu_p(\Phi) = \sup_{\rho \in \mathcal{D}(\mathbb{C}^m)}\|\Phi(\rho)\|_p.
\ee
Here, $\mathcal{D}(\mathbb{C}^m)$ is 
the set of positive semidefinite Hermitian operators of unit trace,
and $\|\;\;\|_p$ is the Schatten $p$-norm: 
$\|A\|_p=({\rm tr}|A|^p)^{\frac{1}{p}}$.

The multiplicativity property was investigated in the context of
quantum information theory.
I.e., 
$\mathcal{D}(\mathbb{C}^m)$ represents quantum states 
on the $m$-dimensional space,
and we restrict the map $\Phi$ in (\ref{MOPN})
to Completely Positive (CP) Trace-Preserving (TP) maps,
which represent quantum channels.
Recall that a map $\Phi$ is CP 
if for any space $\mathbb{C}^d$
the product $\Phi \otimes {\rm 1}_{\mathbb{C}^d}$
is a positive map,
where ${\rm 1}_{\mathbb{C}^d}$ is the identity map 
on $\ma{B}(\mathbb{C}^d)$.
Then,
the following statement,
which is called the multiplicativity of $p$-norm,
 was conjectured in \cite{AHW00}
but was disproved later;
\be \label{mult}  
\nu_p(\Phi \otimes \Omega)= \nu_p(\Phi)\nu_p(\Omega)
\ee
for any $p \in (1, \infty]$ and for all quantum channels $\Phi$ and $\Omega$.
Note that the bound 
$\nu_p(\Phi \otimes \Omega) \geq  \nu_p(\Phi)\nu_p(\Omega)$ 
is straightforward.

The first counterexample, which is called Werner-Holevo channel, 
was found in \cite{WH02} for $p>4.79$ and $m=n=3$.
Then later, the above conjecture was shown to be false for any $p >1$
if we choose large enough $m$ and $n$ 
(the dimension of the input and output spaces)  \cite{HaydenWinter08}.
However when $p=2$, for example,
we still don't know whether or not 
there is a counterexample for (\ref{mult})
of low dimension.
In this paper, we show, in Theorem \ref{main} and Theorem \ref{main2},
that 
for 
any Positive Trace-Preserving (PTP)  map 
$\Phi:  \ma{B}(\mathbb{C}^3) \rightarrow \ma{B}(\mathbb{C}^2)$
and any CP map 
$\Omega: \ma{B}(\mathbb{C}^m) \rightarrow \ma{B}(\mathbb{C}^n)$
\be\label{mult2}
\nu_p(\Phi \otimes \Omega)= \nu_p(\Phi)\nu_p(\Omega)
\ee 
for $p=2$ and $p\geq4$
as a slight generalization of the corresponding result in \cite{KingKoldan06}.
This result is interesting as the Werner-Holevo channel
is a map from $ \ma{B}(\mathbb{C}^3)$ to $ \ma{B}(\mathbb{C}^3)$
violating multiplicativity for $p>4.79$.
There are some general results in
\cite{GLR05},\cite{KNR04},\cite{KR04},
where sufficient conditions for the multiplicativity were derived.
However these sufficient conditions have not been verified in general.

The above conjecture attracted attention in the relation to
the additivity conjecture \cite{KR99}. 
The additivity conjecture was proven to be globally equivalent 
to the additivity of Holevo capacity and 
the additivity of entanglement of formation \cite{Sho03},
however, it was disproved recently \cite{Hastings08}.
Although, the additivity does not hold in general
it is still interesting to look for classes of channels
for which the additivity is true.
For this
the multiplicativity for $p$ close to $1$ 
can be used to
prove the additivity \cite{AHW00}.
Under some conditions,
the multiplicativity for rather large $p$
implies the additivity \cite{WE05}.

\section{Maps to $\ma{B}(\mathbb{C}^2)$}
Suppose that $\rho$ is a Hermitian operator of unit trace on $\mathbb{C}^2$.
Then, there exists ${\bf w} \in {\bf R^3}$ such that
\be 
\rho = \bar{I} 
+ \frac{1}{2} \sum_{k=1}^3 w_k \sigma_k.
\ee

Here, $\bar{I} = I/2$ is the normalized identity and $\sigma_k$ are the Pauli matrices.
Note that $\rho$ is positive semidefinite if and only if 
$\|\rho\|_2 = \|{\bf w}\|_2 \leq 1$, and
$\rho$ is a rank-one projection if and only if
$\|\rho\|_2 =\|{\bf w}\|_2 = 1$.
We identify a quantum state with a vector in the unit ball in $\mathbb{R}^3$.
In this case,
a pure state,
which is a rank-one projection,
 corresponds to a point on the unit sphere.
This unit ball is called the Bloch ball, denoted by $B_1$.
Note that the center corresponds to the maximally mixed state.
The following estimate is also important.
\be
\| \rho \|_2 = 
\sqrt{\frac{1}{2} + \frac{1}{2} \sum_{k=1}^3 w_k^2}.
\ee
Note that the 2-norm is determined by the distance from the center
and then
this fact shows that $\nu_2(\Phi)$ is also determined by the minimum radius of ball
which includes $\Phi(B_1)$ the image of the Bloch ball by $\Phi$.
This observation can be extended to $p \in (1, \infty]$
by using the majorization of eigenvalues.
 
The depolarizing channel on $\ma{B}(\mathbb{C}^d)$ is defined as
\be
\Psi_{\lambda}(\rho) = \lambda \rho + (1-\lambda){\rm tr} [\rho] \bar{I}.
\ee
Here, $\bar{I} = I/d$ and $0\leq \lambda \leq 1$.
Then, when $d=2$ it acts on the above quantum states as follows.
\be
\Psi_\lambda (\rho) =  \bar{I} + \frac{1}{2}\sum_{k=1}^3 \lambda \, w_k \sigma_k. 
\ee
The depolarizing channel $\Psi_\lambda$ compresses $B_1$
to the ball with radius $\lambda$,
which is denoted by $B_\lambda$.

\begin{thm}\label{m-dec}
Any PTP map 
$\Phi : \ma{B}(\mathbb{C}^n) \rightarrow \ma{B}(\mathbb{C}^2) $ can be written in the form 
\footnote{
This form of decomposition may be traced back to our previous paper \cite{Fuk05}.
}
of
\be
\Phi = \Psi_\lambda \circ M.
\ee
Here, $\Psi_\lambda$ is the depolarizing channel on $\ma{B}(\mathbb{C}^2)$ 
and $M : \ma{B}(\mathbb{C}^n) \rightarrow \ma{B}(\mathbb{C}^2)$ 
is a PTP map which has a rank-one-projection output,
so that
\be
\nu_p(\Phi) = \nu_p(\Psi_\lambda)
\quad p\in (1,\infty].
\ee
\end{thm}
{\bf Proof.}
First, recall that 
the depolarizing channel on $\ma{B}(\mathbb{C}^2)$
is defined by the following mappings.
\be
&\Psi_\lambda : \ma{B}(\mathbb{C}^2)\rightarrow \ma{B}(\mathbb{C}^2)\notag\\
&I \mapsto I;\qquad
\sigma_1 \mapsto \lambda \sigma_1;\qquad
\sigma_2 \mapsto \lambda \sigma_2;\qquad
\sigma_3 \mapsto \lambda \sigma_3.
\ee
We define a new map for $0<\lambda\leq 1$:
\be
&L_{\lambda} : \ma{B}(\mathbb{C}^2)\rightarrow \ma{B}(\mathbb{C}^2)\notag\\
&I \mapsto I;\qquad
\sigma_1 \mapsto \frac{1}{\lambda} \sigma_1;\qquad
\sigma_2 \mapsto \frac{1}{\lambda} \sigma_2;\qquad
\sigma_3 \mapsto \frac{1}{\lambda} \sigma_3.
\ee

Then, next, choose $0 \leq \lambda \leq 1$ such that $\nu_p(\Phi)=\nu_p(\Psi_\lambda )$.
Since when $\lambda = 0$
($\Phi$ has only one output $\bar{I}$ and $\nu_2(\Phi)=1/\sqrt{2}$)
the statement of theorem holds,
we assume that $\lambda > 0$.
Then $L_{\lambda}$ is well-defined and the channel $\Phi$ can be written as
\be
\Phi= \Psi_\lambda \circ L_{\lambda} \circ \Phi.
\ee
Here, $\Psi_\lambda \circ L_{\lambda}$ acts as the identity.

Finally, we show the map $M=L_{\lambda} \circ \Phi$ is PTP 
and has a rank-one-projection output.
Note that a TP map $M$ is positive iff $M(B_1)\subseteq B_1$.
The condition $\nu_p(\Phi)=\nu_p(\Psi_\lambda )$ implies that
$\Phi(B_1)$ is touching $B_\lambda$ from the inside.
Hence, 
\be
M(B_1)= L_{\lambda} ( \Phi (B_1))
\subseteq L_{\lambda} (B_\lambda) = B_1.
\ee
This shows that the map $M$ is positive and that
 $M(B_1)$
is touching $B_1$ from inside so that 
$M$ has a rank-one-projection output.
By the construction $M$ preserves trace.\\
Q.E.D.

Also, the following result on the depolarizing channels is well-known
\cite{Kin02},\cite{Kin03}.
\begin{thm}\label{depolarizing}
Let $\Psi_\lambda$ be the depolarizing channel.
Then, $\nu_p (\Psi_\lambda \otimes \Omega)
\leq \nu_p (\Psi_\lambda)\,\nu_p (\Omega)$  
\end{thm}
for any CP map $\Omega$ and $p \in (1, \infty]$.

\section{Decomposability and its application}
In this section, 
we use the concept of decomposability to prove multiplicativity properties 
for PTP maps between low dimensional spaces.

\begin{defn}
A positive map $M$ is decomposable if
\be
M= \Phi_1 +  T \circ \Phi_2
\ee
for some CP maps $\Phi_1$ and $\Phi_2$.
Here, $T$ is the transpose map.
\end{defn}

The following result is well-known \cite{Stormer63},\cite{Woronowicz76} and
our result totally depends on it.
\begin{thm}\label{deco}
All positive maps $M: \ma{B}(\mathbb{C}^3)\rightarrow \ma{B}(\mathbb{C}^2)$
and
$M: \ma{B}(\mathbb{C}^2)\rightarrow \ma{B}(\mathbb{C}^3)$
are decomposable.
\end{thm}

Then, we have
\begin{lemma}\label{lemma}
Let $\Phi$ be a PTP map from $\ma{B}(\mathbb{C}^3)$ to $\ma{B}(\mathbb{C}^2)$.
Then, 
\be
\Phi= \Psi_\lambda \circ \Phi_1
 + T \circ  \Psi_\lambda\circ \Phi_2
\ee
for some CP maps $\Phi_1$ and $\Phi_2$,
so that $\nu_p (\Phi) = \nu_p(\Psi_\lambda)$ for $p \in (1,\infty]$.
\end{lemma}
{\bf Proof.}
By Theorem \ref{m-dec} and Theorem \ref{deco}
\be
\Phi &=& \Psi_\lambda \circ M 
= \Psi_\lambda \circ [\Phi_1 + T \circ \Phi_2 ]\notag\\
&=& \Psi_\lambda \circ \Phi_1 + \Psi_\lambda\circ T\circ \Phi_2 
= \Psi_\lambda \circ \Phi_1 + T\circ \Psi_\lambda\circ \Phi_2.
\ee
Note that $\Psi_\lambda$ and $T$ are commutative.
\\
Q.E.D.

\subsection{For $p=2$}
When $p=2$ we have 
the following nice property on the $2$-norm:
\begin{lemma}\label{Tand2norm}
\be
\| \hat{A} \|_2 = \|(T \otimes {\bf 1}_{\mathbb{C}^n}) (\hat{A} )  \|_2
\ee
for any $\hat{A}\in \ma{B}(\mathbb{C}^{mn})$.
\end{lemma}
{\bf Proof.}
$\hat{A}\in \ma{B}(\mathbb{C}^{mn})$ can be written as
\be
\hat{A} = \sum_{i,j=1}^m |i \rangle\langle j| \otimes A_{ij}
\ee
Here, $\{|i\rangle\}$ is an orthonormal basis
and $A_{ij} \in \ma{B}(\mathbb{C}^n)$.
Then,
\be
(T \otimes {\bf 1}_{\mathbb{C}^n}) (\hat{A} )
= \sum_{i,j=1}^m |j\rangle\langle i| \otimes A_{ij}.
\ee
Here, the transpose $T$ is defined in the basis $\{|i\rangle\}$.
Therefore,
\be
\| \hat{A} \|_2^2
=\sum_{i,j=1}^m \| A_{ij} \|_2^2
= \|(T \otimes {\bf 1}_{\mathbb{C}^n}) (\hat{A} )  \|_2^2.
\ee
Q.E.D.

\begin{thm} \label{main}
Let $\Phi$ be a PTP map from $\ma{B}(\mathbb{C}^3)$ to $\ma{B}(\mathbb{C}^2)$.
Then, for any CP map 
$\Omega: \ma{B}(\mathbb{C}^m)\rightarrow \ma{B}(\mathbb{C}^n)$,
\be
\nu_2 (\Phi \otimes \Omega) = \nu_2 (\Phi) \, \nu_2 (\Omega).
\ee
\end{thm}
{\bf Proof.}
We show $\nu_2 (\Phi \otimes \Omega) \leq \nu_2 (\Phi) \, \nu_2 (\Omega)$
as the other inequality is obvious.

For any state $\hat{\rho} \in \ma{D}(\mathbb{C}^3 \otimes \mathbb{C}^m )$
let $\sigma_1$ and $\sigma_2$ be positive semidefinite Hermitian operators as follows;
\be
\sigma_1 = (\Phi_1 \otimes {\bf 1} )(\hat{\rho})
\; \text{and}\;
\sigma_2 = (\Phi_2 \otimes {\bf 1} )(\hat{\rho}).
\ee
Here, $\Phi_1$ and $\Phi_2$ are as in Lemma \ref{lemma}.
Then, 
\be \label{deco2}
(\Phi \otimes {\bf 1})(\hat{\rho}) 
= (\Psi_\lambda \otimes {\bf 1} )( \sigma_1) 
+  ((T \circ \Psi_\lambda )\otimes {\bf 1} )( \sigma_2)
\ee
Also, since $\Phi$, $\Psi_\lambda$ and $T$ preserve trace, 
\be \label{convex}
1={\rm tr}[(\Phi \otimes {\bf 1} )(\hat{\rho})]
= {\rm tr}[\sigma_1] 
+ {\rm tr}[\sigma_2]. 
\ee

Next, Theorem \ref{depolarizing} gives the following bounds.
\be
\| (\Psi_\lambda \otimes \Omega )( \sigma_1) \|_2 
\leq \nu_2(\Psi_\lambda)\, \nu_2(\Omega)\, {\rm tr}[\sigma_1]
\quad \text{and} \quad
\| (\Psi_\lambda \otimes \Omega )( \sigma_2) \|_2 
\leq \nu_2(\Psi_\lambda)\, \nu_2(\Omega)\, {\rm tr}[\sigma_2]
\label{twobounds}\ee
Then, 
by using (\ref{deco2}), the triangle inequality,
Lemma \ref{Tand2norm}, (\ref{twobounds}) and
 (\ref{convex}) in order,
\be 
\| (\Phi \otimes \Omega) (\hat{\rho})  \|_2 
&\leq&  \| (\Psi_\lambda \otimes \Omega )( \sigma_1) \|_2 
+ \|(T\otimes {\bf 1})\circ (\Psi_\lambda \otimes \Omega )( \sigma_2)\|_2\notag\\
&=& \| (\Psi_\lambda \otimes \Omega )( \sigma_1) \|_2 
+ \|(\Psi_\lambda \otimes \Omega )( \sigma_2)\|_2\notag\\
&\leq & \nu_2(\Psi_\lambda)\, \nu_2(\Omega) 
\left[ {\rm tr}[\sigma_1] 
+ {\rm tr}[\sigma_2] \right]\notag\\
& = & \nu_2(\Phi)\, \nu_2(\Omega).
\ee
This implies that
\be
\nu_2(\Phi \otimes \Omega) \leq \nu_2(\Phi)\, \nu_2(\Omega).
\ee
Q.E.D.
 
\subsection{For $p \geq 4$} 

To get the result for $p \geq 4$ we need the following result \cite{KingKoldan06}.
Note that it is also possible to use Theorem \ref{KK}
instead of Lemma \ref{Tand2norm}
to prove Theorem \ref{main}.
\begin{thm}\label{KK}
Let $A, B, C, D \in \ma{B}(\mathbb{C}^d)$ for $d \geq 1$. Then,
\be
\left\| 
\begin{pmatrix}
A&B\\C&D
\end{pmatrix}
\right\|_p \leq 
\left\| 
\begin{pmatrix}
\|A\|_p &\|B\|_p \\ 
\|C\|_p&\|D\|_p
\end{pmatrix}
\right\|_p 
\ee
for $p=2$ and $p \geq 4$.
\end{thm}

\begin{thm}\label{main2}
Let $\Phi$ be a PTP map from $\ma{B}(\mathbb{C}^3)$ to $\ma{B}(\mathbb{C}^2)$.
Then, for any CP map 
$\Omega: \ma{B}(\mathbb{C}^m)\rightarrow \ma{B}(\mathbb{C}^n)$,
\be
\nu_p (\Phi \otimes \Omega) = \nu_p (\Phi) \, \nu_p (\Omega).
\ee
for $p\geq4$.
\end{thm}
{\bf Proof.} We can prove the above statement in a similar way as Theorem \ref{main}.
One step which is not trivial is the following bound:
\be
\|(T\otimes {\bf 1})\circ (\Psi_\lambda \otimes \Omega )( \sigma_2)\|_p
\leq \nu_p (\Psi_\lambda)\, \nu_p(\Omega) \,
{\rm tr}[\sigma_2].
\ee
Here, we use the same notations as in the proof of Theorem \ref{main}.
To get this bound write
\be
\sigma_2 = 
\begin{pmatrix}
A&B\\B^\ast &C
\end{pmatrix}
\ee
for some $A,B,C \in \ma{B}(\mathbb{C}^m)$.
Note that since $\sigma_2$ is positive semidefinite, so are $A$ and $C$.
Then,
\be
\|(T\otimes {\bf 1})\circ (\Psi_\lambda \otimes \Omega )( \sigma_2)\|_p
&=
\left\|
\begin{pmatrix}
\frac{1+\lambda}{2}\Omega(A) + \frac{1-\lambda}{2} \Omega(C)& \lambda \Omega(B^\ast)\\
\lambda \Omega(B) & \frac{1-\lambda}{2}\Omega(A) + \frac{1+\lambda}{2} \Omega(C)
\end{pmatrix}
\right\|_p 
\ee
By Theorem \ref{KK} and the triangle inequality, it is bounded by
\be
&& \left\|
\begin{pmatrix}
\frac{1+\lambda}{2}\|\Omega(A)\|_p + \frac{1-\lambda}{2} \|\Omega(C)\|_p
& \lambda \|\Omega(B^\ast)\|_p\\
\lambda \|\Omega(B)\|_p 
& \frac{1-\lambda}{2} \|\Omega(A)\|_p + \frac{1+\lambda}{2} \|\Omega(C)\|_p
\end{pmatrix} 
\right\|_p \notag\\
&=& \left\|
\Psi_\lambda 
\left(
\begin{pmatrix}
\|\Omega(A)\|_p 
&\|\Omega(B^\ast)\|_p\\
\|\Omega(B)\|_p 
&  \|\Omega(C)\|_p
\end{pmatrix} 
\right)
\right\|_p \notag\\
&\leq & \nu_p (\Psi_\lambda)  \,
\left[\|\Omega(A)\|_p + \|\Omega(C)\|_p \right] \notag\\
&\leq &
\nu_p (\Psi_\lambda) \, \nu_p (\Omega) \,[{\rm tr} [A] + {\rm tr} [C] ]\notag\\
&=& \nu_p (\Psi_\lambda) \, \nu_p (\Omega) \,{\rm tr}[\sigma_2].
\ee
Here, we used the fact that the following $2 \times 2$ matrices 
\be
\begin{pmatrix}
\|\frac{1+\lambda}{2}\Omega(A) + \frac{1-\lambda}{2} \Omega(C)\|_p& 
\|\lambda \Omega(B^\ast)\|_p \\
\|\lambda \Omega(B)\|_p & 
\|\frac{1-\lambda}{2}\Omega(A) + \frac{1+\lambda}{2} \Omega(C)\|_p
\end{pmatrix}
\; \text{and}\;
\begin{pmatrix}
\|\Omega(A)\|_p 
&\|\Omega(B^\ast)\|_p\\
\|\Omega(B)\|_p 
&  \|\Omega(C)\|_p
\end{pmatrix}
\label{positivematrix}
\ee
are positive semidefinite.
Indeed,
since 
\be
\begin{pmatrix}
\Omega(A) 
&\Omega(B)\\
\Omega(B^\ast) 
&  \Omega(C)
\end{pmatrix}
\ee
is positive semidefinite we can write
$\Omega(B) = \Omega(A)^{1/2} \, R \, \Omega(C)^{1/2}$
for some contraction $R$ but this gives the bound:
$\|\Omega(B)\|_p \leq \sqrt{\|\Omega(A)\|_p \| \Omega(C) \|_p}$
and hence the positivity in (\ref{positivematrix}).

Since the following bound:
\be
\|(\Psi_\lambda \otimes \Omega )( \sigma_1)\|_p
\leq \nu_p (\Psi_\lambda)\, \nu_p(\Omega) \,
{\rm tr}[\sigma_1]
\ee
is derived in a similar way we have
\be 
\| (\Phi \otimes \Omega) (\hat{\rho})  \|_p
\leq \nu_p(\Phi)\, \nu_p(\Omega).
\ee
Q.E.D.

\begin{rmk}{\rm
We take $\Omega$ as a CP map but the $2$-positivity is sufficient.
A similar observation holds in the following section as well.
}
\end{rmk}

\subsection{Generalization and corollaries} 
 
Any CP map $\Phi$ from $\ma{B}(\mathbb{C}^m)$ to $\ma{B}(\mathbb{C}^n)$
can be written in the Kraus form:
\be\label{Kraus}
\Phi(\rho) =\sum_{k=1}^N A_k \rho A_k^\ast.
\ee
Here, $A_k$ are $n \times m$ matrices.
The condition $\sum_{k=1}^N A_k^\ast A_k =I$
implies that $\Phi$ is TP.
We also define the complementary/conjugate channel of $\Phi$ as follows.
\be
\Phi^C(\rho) = {\rm tr} [A_k \rho A_l^\ast] |k\rangle \langle l|.  
\ee 
Note that this is a CPTP map 
from $\ma{B}(\mathbb{C}^m)$ to $\ma{B}(\mathbb{C}^N)$,
whose dimension is the number of Kraus operators in (\ref{Kraus}).
As in \cite{Hol05}, \cite{KMNR05},
a channel and its complementary/conjugate channel share 
the maximal output $p$-norm and then the multiplicativity property.
Therefore,
Theorem \ref{main} and Theorem \ref{main2}
give the following corollary.
\begin{cor}\label{3to3}
Let $\Phi$ be a CPTP map from $\ma{B}(\mathbb{C}^3)$ to $\ma{B}(\mathbb{C}^n)$.
If $\Phi$ can be written by two Kraus operators then
$ \nu_p(\Phi \otimes \Omega) = \nu_p(\Phi)\, \nu_p(\Omega)$
for $p=2$ and $p \geq 4$.
\end{cor}
{\bf Proof.}
$\Phi^C$ is a CPTP map from $\ma{B}(\mathbb{C}^3)$ to $\ma{B}(\mathbb{C}^2)$.
Hence, by using Theorem \ref{main} and Theorem \ref{main2}, the statement follows.\\
Q.E.D.

Also, we can generalize Theorem \ref{main}:
\begin{thm}\label{general}
Suppose we have a PTP map $\Phi = \Psi_\lambda \circ M$.
Here, $M$ is a PTP decomposable map from 
$\ma{B}(\mathbb{C}^m)$ to $\ma{B}(\mathbb{C}^n)$  
having a rank-one-projection output,
and 
$ \Psi_\lambda$ is the depolarizing channel on $\ma{B}(\mathbb{C}^n)$.
Then $\nu_2(\Phi\otimes \Omega) = \nu_2(\Phi)\, \nu_2(\Omega)$
for any CP map $\Omega$.
\end{thm} 
The above statement can be proven in a similar way as Theorem \ref{main}, and
it is a generalization of the result in \cite{Fuk05} 
when $p=2$.
Note that this statement is not vacuous.
For example, take two CPTP maps $\Phi_1$ and $\Phi_2$ such that
$\Phi_1$ and $T \circ \Phi_2$ have the common rank-one-projection output.
Then, $M= q\,\Phi_1 + (1-q) \, T \circ \Phi_2$
for $0\leq q \leq 1$
satisfies the above condition.

\begin{cor}
Suppose we have a PTP map $\Phi = \Psi_\lambda \circ M$.
Here, $M$ is a PTP map from 
$\ma{B}(\mathbb{C}^2)$ to $\ma{B}(\mathbb{C}^3)$ 
having a rank-one-projection output,
and 
$ \Psi_\lambda$ is the depolarizing channel on $\ma{B}(\mathbb{C}^3)$.
Then $\nu_2(\Phi\otimes \Omega) = \nu_2(\Phi)\, \nu_2(\Omega)$
for any CP map $\Omega$.
\end{cor}
{\bf Proof.}
By Theorem \ref{deco}, $M$ is always decomposable.
Hence by Theorem \ref{general} the result follows.\\
Q.E.D. 

\section{Discussion}

In this paper, 
we used the concept of decomposability of positive maps.
Since partial transpose does not preserve positivity
we had to exclude the case $p \in (2,4)$.
It would be interesting to investigate
whether or not the same bound holds for $p \in (2,4)$.
There is another interesting question.
We don't know very much about decomposability of positive maps
$M: \ma{B}(\mathbb{C}^m) \rightarrow \ma{B}(\mathbb{C}^2)$
when $m>3$ although some researches are being done 
\cite{MajewskiMarciniak05}.
Decomposable maps of this class will give other 
PTP maps which have multiplicativity property.

\begin{center}
{\bf Acknowledgment }
\end{center}
M.F. thanks Bruno Nachtergaele for supporting this research. 
A. Holevo is thanked for giving a useful comment on the draft.
The result for $p=2$ was presented at the workshop held 
at University of Arizona in June 2009,
and M.F. thanks Robert Sims and other people who organized it.
Christopher King is thanked for 
letting me know after the presentation
that the original result on $p=2$
can be extended to $p \geq 4$, and 
that Corollary \ref{3to3} holds.

\end{document}